\documentclass[%
 reprint,
 amsmath,amssymb,
 aps,
 prc,
]{revtex4-1}

\usepackage{braket}
\usepackage{graphicx}
\usepackage{dcolumn}
\usepackage{bm}
\usepackage{url}
\usepackage{hyperref}
\usepackage{threeparttable}
\usepackage{mathtools}
\usepackage[capitalise]{cleveref}
\usepackage{tabularx}
\usepackage{physics}

\begin{document}


\title{Convergence of electric quadrupole rotational invariants from the nuclear shell model}

\author{J.~Henderson}
\email{jack.henderson@surrey.ac.uk}
\altaffiliation{Department of Physics, University of Surrey, Guildford, GU2 7XH, United Kingdom}
\affiliation{Lawrence Livermore National Laboratory, Livermore, CA 94550, USA}

\date{\today}

\begin{abstract}
\begin{description}
\item[Background] Nuclei exhibit both single-particle and collective degrees of freedom, with the latter often subdivided into vibrational and rotational motions. Experimentally identifying the relative roles of these collective modes is extremely challenging, particularly in the face of possible shape coexistence.
\item[Purpose] Model-independent, invariant quantities describing the deformation of a nucleus in the intrinsic frame have long been known but their determination potentially requires a large quantity of experimental data to achieve convergence. Through comparison with the nuclear shell model, the question of convergence will be addressed.
\item[Methods] Shell-model calculations performed in the $sd$- and $pf$-shell model spaces are used to determine electric-quadrupole matrix elements for a multitude of low-lying states using the first forty states of the relevant spins. Relative contributions to the rotationally invariant quantities from multiple states can therefore be determined.
\item[Results] It is found that on average, the inclusion of four intermediate states results in the leading-order invariant, $\left<\hat{Q^2}\right>$, converging to within 10\% of its true value and the triaxiality term, $\cos{\left(3\delta\right)}$, converging to its true value, though some variance remains. Higher-order quantities relating to the softness of the nuclear shape are found to converge more slowly.
\item[Conclusions] The convergence of quadrupole rotationally invariant sum rules was quantified in the $sd$- and $pf$-shell model spaces and indicates the challenge inherent in a full determination of nuclear shape. The present study is limited to relatively small valence spaces. Larger spaces, such as the rare-earth region, potentially offer faster convergence.
\end{description}
\end{abstract}

\pacs{Valid PACS appear here}
\maketitle

\section{Introduction}

\begin{figure*}
\begin{minipage}[h]{0.49\linewidth}
\centerline{\includegraphics[width=\linewidth]{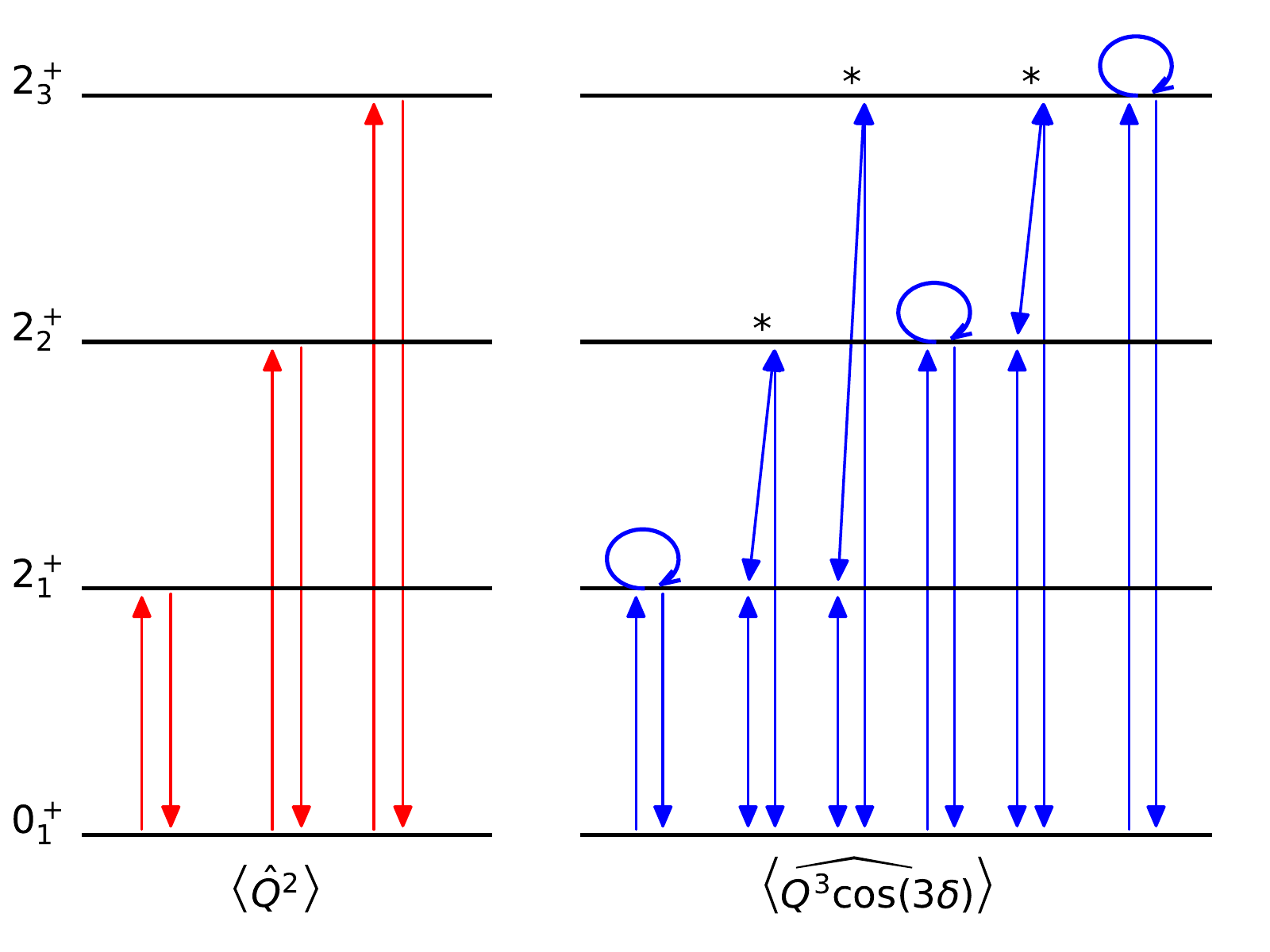}}
\end{minipage}
\begin{minipage}[h]{0.49\linewidth}
\centerline{\includegraphics[width=\linewidth]{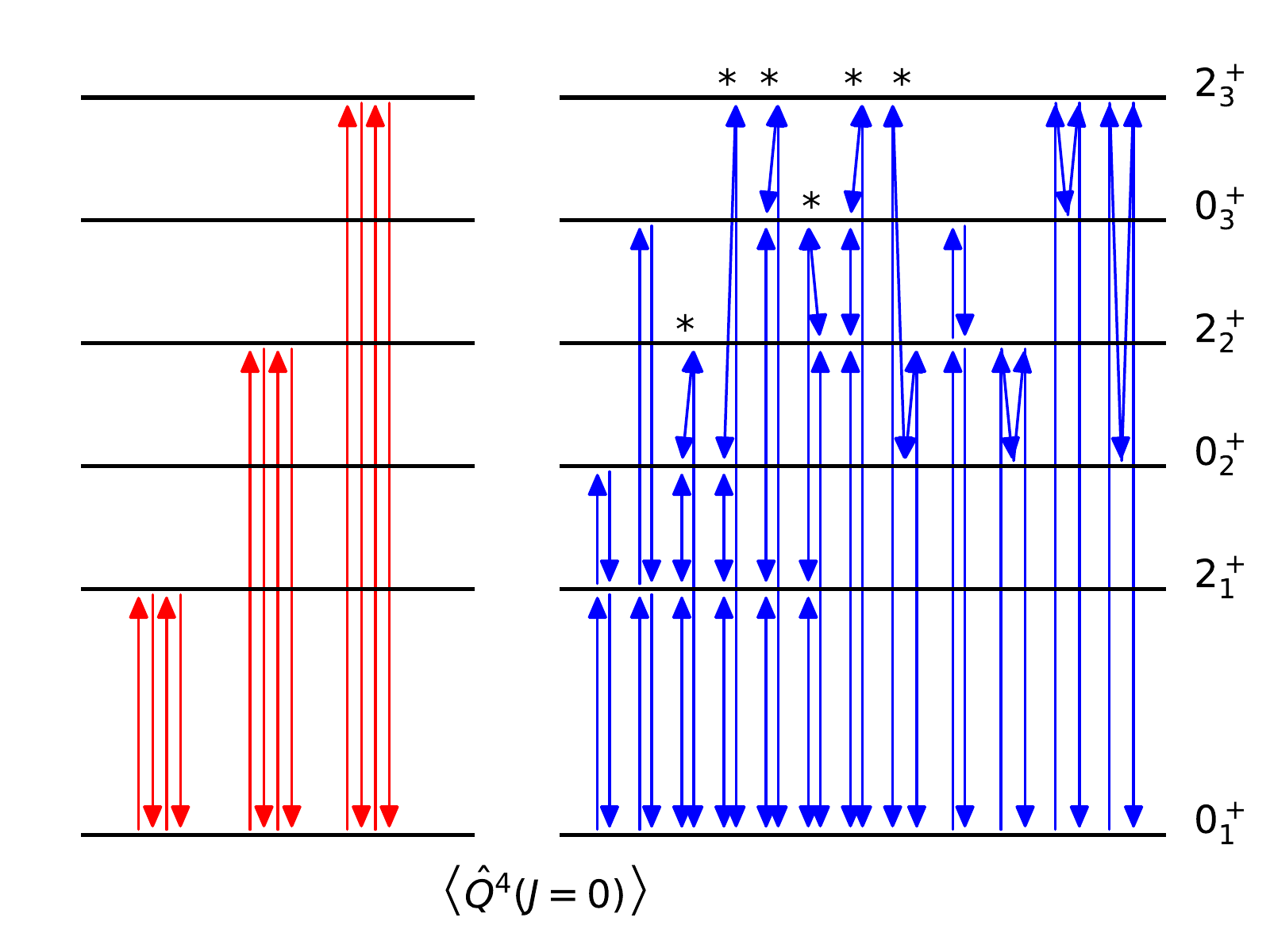}}
\end{minipage}
\caption{Electric quadrupole matrix elements contributing to the first three rotationally invariant expectation values for the $0^+$ ground state. $\ev{\hat{Q^2}}$, $\ev{\widehat{Q^3\cos{3\delta}}}$ and $\ev{\hat{Q^4}}$ are shown for a simplified level scheme containing only the first three states of relevant $J^\pi$. Matrix elements for $\ev{\hat{Q^4}}$ are separated into those which contribute to $\ev{\hat{Q^2}}$ (left, red) and further loops (right, blue). The significant increase in the number of matrix elements required for $\ev{\hat{Q^4}}$ over those required for $\ev{\hat{Q^2}}$ and $\ev{\widehat{Q^3\cos{3\delta}}}$ is clearly apparent. Asymmetric loops, indicated by *, contribute twice to the sum.}
\label{fig:loops}
\end{figure*}

Atomic nuclei exhibit properties associated with the collective motion of their constituent nucleons arising from quadrupole deformation. As a phenomenon which necessarily involves a large number of nucleons, understanding collective behaviour, and therefore the deformation of the nucleus, presents an exceptional challenge for microscopic nuclear models. In spite of the well-established collective behaviour of nuclei, the signatures of collective motion are often complex, with the disentangling of collective rotations and vibrations, and the motion of single particles within the nucleus proving a longstanding challenge. Key to understanding collective behavior is identifying relevant experimental observables and assessing their relation to the phenomenon. 

One powerful experimental method utilises quadrupole rotationally invariant sum rules~\cite{ref:Kumar_72,ref:Cline_86}, which provide experimental access to the nucleus's quadrupole deformation in a model-independent manner and have been widely employed experimentally (see e.g. Refs.~\cite{ref:Ayangeakaa_19,ref:Henderson_19, ref:Ayangeakaa_16, ref:Hadynska_16, ref:Wrzosek_12, ref:Clement_07, ref:Srebny_06, ref:Wu_96}), and theoretically (e.g. Refs~\cite{ref:Poves_19,ref:Gilbreth_18,ref:Naidja_17,ref:Quan_17,ref:Schmidt_17}). In this work, shell-model calculations will be used to assess the convergence of these sum rules. It is noted that the theoretical method presented here differs from that given in Refs.~\cite{ref:Poves_19,ref:Gilbreth_18} which avoid the issue of convergence, and which address the deformation of the full nuclear matter (protons and neutrons). Here, the goal is to inform experiment through an investigation of the electric quadrupole sum-rule convergence and so the standard experimental method is employed. The discussion will be prefaced with mention of more widely used observables used to characterise deformation.
  
\subsection{Rotational invariants}

A number of experimental signatures have been used to quantify nuclear deformation and the associated property of collectivity. The energy of the first excited $2^+$ state in an even-even nucleus is often used to infer the degree of collectivity, and therefore increased deformation. From geometric arguments, one can also determine the form of collectivity (vibrational or rotational) from the ratio of $4^+_1$ and $2^+_1$ energies, for example. Beyond excitation energies, large electric quadrupole transitions strengths ($B(E2)$ values) indicate enhanced quadrupole collectivity and have been related~\cite{ref:Pritychenko_16} to the magnitude of quadrupole deformation, defined by the $\beta_2$ parameter of the Bohr Hamiltonian

\begin{equation}
\beta_2 = \frac{4\pi}{3ZR_0^2}\sqrt{\frac{B(E2;0^+_1\rightarrow2^+_1)}{e^2}}.
\label{eq:beta_2}
\end{equation}

Here, $R_0=1.2A^{1/3}$~fm and the $B(E2)$ is in units of $e^2fm^4$, where the assumption is that the charge- and matter-distributions are the same. Spectroscopic quadrupole moments, $Q_s(I)$, can be used to infer the prolate or oblate nature of the nuclear deformation. Nuclei are not limited to axially symmetric rotational and vibrational structures, however, and commonly assume triaxial shapes. One can estimate the role of non-axial deformation in a nucleus from the ratio of the experimentally determined $Q_s(2^+_1)$ and that expected from an axial rotor, based on the $B(E2;0^+_1\rightarrow2^+_1)$ value. Assuming axial symmetry, 

\begin{equation}
\left| Q_s(2^+_1) \right| = \frac{2}{7}\sqrt{\frac{16\pi}{5} \cdot B(E2;0^+_1\rightarrow2^+_1)}.
\end{equation}

A $Q_s(2^+_1)$ that is smaller than expected is often indicative of triaxiality, with a maximally triaxial system ($\gamma = 30^\circ$ in the terminology of the Bohr Hamiltonian) yielding $Q_s(2^+_1)=0$ in a rigid asymmetric rotor~\cite{ref:Davydov_58}, for example, regardless of the $B(E2;2^+_1\rightarrow0^+_1)$ value.

While the above signatures can be used to provide a first indication of the nuclear deformation, the work of Kumar~\cite{ref:Kumar_72} and Cline~\cite{ref:Cline_86} provides a model-independent method to quantify the nuclear shape and its softness. Electromagnetic multipole operators are spherical tensors and zero-coupled products of the operators are therefore rotationally invariant, providing a method by which one can relate quantities in the laboratory and intrinsic nuclear frames. Utilising an intermediate state expansion, these zero-coupled operator products can take the form of sums of products of $E2$ matrix elements arranged in loops, as shown in Fig.~\ref{fig:loops} in a simplified form.

In this work, the Cline notation will be used, as outlined in Refs.~\cite{ref:Cline_86,ref:GOSIA_manual}. Parameters denoted $Q$ and $\delta$ are used to define the charge distribution in the intrinsic frame with regards to the electric multipole operator $E(\lambda,\mu)$:

\begin{align}
& E(2,0) = Q\cos{\left(\delta\right)} \\
& E(2,1) = E(2,-1) = 0 \\
& E(2,2) = E(2,-2) = \frac{1}{\sqrt{2}}Q\sin{\left(\delta\right)}.
\end{align}

These can be thought of in analogy to the Bohr shape parameters ($\beta,\gamma$) which define the radial shape of a quadrupole-deformed object. The shorthand notation

\begin{equation}
M_{if} = \mel{i}{\hat{E2}}{f}
\end{equation}

\noindent is employed for the reduced matrix elements, where $\hat{E2}$ is the electric quadrupole operator and $i$ and $f$ correspond to the initial and final states, respectively. In the present work, $I$ is used to refer to the spin of a given state, while $J$ refers to the angular momentum coupling within the rotational invariants. Here, for completeness, the invariant definitions provided in Ref.~\cite{ref:GOSIA_manual} are reproduced. The first invariant yields the expectation value for $Q^2$:

\begin{equation}
\ev{\hat{Q^2}} = \sqrt{5}\ev{[\hat{E2}\times \hat{E2}]_0}{s}
\end{equation}

while the second gives the expectation value

\begin{equation}
\ev{\widehat{Q^3\cos{\left(3\delta\right)}}} = -\frac{\sqrt{35}}{\sqrt{2}}\ev{\{[\hat{E2}\times \hat{E2}]_2\times \hat{E2}\}_0}{s},
\end{equation}

where $s$ is the state of interest. Note that the notations on the left hand sides of the above equations (and similar notations throughout this work) use operators that are symbolically denoted by their eigenvalues in the intrinsic frame. Using an intermediate-state expansion:

\begin{equation}
\ev{[\hat{E2}\times \hat{E2}]_0}{s} = \frac{(-1)^{2I_s}}{\sqrt{\left(2I_s+1\right)}}\sum_t M_{st}M_{ts}\left\{\begin{matrix}2 & 2 & 0 \\ I_s & I_s & I_t\end{matrix}\right\}
\label{eq:Q2}
\end{equation}

and

\begin{multline}
\ev{\{[\hat{E2}\times \hat{E2}]_2\times \hat{E2}\}_0}{s} =\\ 
	\frac{(-1)^{2I_s}}{2I_s+1}\sum_{tu} M_{su}M_{ut}M_{ts}\left\{\begin{matrix}2 & 2 & 2 \\ I_s & I_t & I_u\end{matrix}\right\}.
\label{eq:Q3cos3d}
\end{multline}

Here, $\left\{\begin{matrix} ... \end{matrix}\right\}$ correspond to Wigner-6j symbols. Based on the above one can determine the expectation value for the absolute degree of deformation, $\ev{\hat{Q^2}}$ and, as in e.g. Ref.~\cite{ref:Alhassid_14}, extract the triaxiality parameter, $\delta$, from

\begin{align}
\cos{(3\delta)} = -\frac{\sqrt{7}}{\sqrt{2}\sqrt[4]{5}}\frac{\ev{\{[\hat{E2}\times \hat{E2}]_2\times \hat{E2}\}_0}{s}}{\ev{[\hat{E2}\times \hat{E2}]_0}{s}^{3/2}} \nonumber \\ = \frac{\ev{\widehat{Q^3\cos{\left(3\delta\right)}}}}{\ev{\hat{Q^2}}^{3/2}}.
\label{eq:cos}
\end{align}

Note that $\mel{0^+_1}{\hat{E2}}{2^+_1}$ typically has the largest magnitude of those $E2$ matrix elements connecting to the ground state. This allows one to investigate parallels to the aforementioned experimental signatures by including only the $0^+_1$ and $2^+_1$ states and to define approximations of the two above invariant quantities for the $0^+_1$ ground state in an even-even nucleus. With this approximation:

\begin{equation}
\ev{\hat{Q^2}} \approx B(E2;0^+_1\rightarrow2^+_1)
\label{eq:q2_LO}
\end{equation}

and,

\begin{equation}
\cos{(3\delta)} \approx -\frac{Q_s(2^+_1)}{\frac{2}{7}\sqrt{\frac{16\pi}{5} \cdot B(E2;0^+_1\rightarrow2^+_1)}}.
\label{eq:cos_approx}
\end{equation}

The standard parameterisation of the $\beta_2$ deformation parameter given in Eq.~\ref{eq:beta_2} can be clearly identified as relating to the expectation value $\ev{\hat{Q^2}}$ definition approximated in Eq.~\ref{eq:q2_LO}. Note that these approximate solutions correspond to the leftmost $\hat{E2}$ operator product ``loops'' in the $\ev{\hat{Q^2}}$ and $\ev{\widehat{Q^3\cos{\left(3\delta\right)}}}$ panels of Fig.~\ref{fig:loops}. 

Higher-order invariant quantities can also be constructed, but require different intermediate $J$-couplings of the $\hat{E2}$ operator products. Here for brevity, only the $J=0$ couplings will be provided. Other couplings are given in Ref.~\cite{ref:GOSIA_manual}. One can write down an expectation value arising from the fourth product, 

\begin{equation}
\ev{{\hat{P^4}}(J)} = \mel{s}{\{(\hat{E2}\times \hat{E2})_J \times (\hat{E2}\times \hat{E2})_J\}_0}{s},
\end{equation}

where

\begin{multline}
\bra{s}\left\{\left(\hat{E2}\times \hat{E2}\right)_J \times \left(\hat{E2}\times \hat{E2}\right)_J\right\}_0\ket{s}= \\
	\frac{\sqrt{2J+1}}{2I_s + 1} \sum_{rtu}M_{st}M_{tr}M_{ru}M_{us} \\
	\left\{\begin{matrix} 2 & 2 & J \\ I_s & I_r & I_t \end{matrix}\right\}\left\{\begin{matrix} 2 & 2 & J \\ I_s & I_r & I_u \end{matrix}\right\} (-1)^{I_s - I_r},
\end{multline}

and yields the expectation value

\begin{equation}
\left<\hat{Q^4}(J=0)\right> = 5\left<\hat{P^4}(0)\right>.
\end{equation}

Using the expectation values $\ev{\hat{Q^4}}$ and $\ev{\hat{Q^2}}$ and relating the root-mean-square of $Q^2$ one can produce a parameter which defines the width (softness) of the $Q^2$ quantity:

\begin{equation}
\sigma\left(Q^2\right) = \sqrt{\left<\hat{Q^4}\right> - \left<\hat{Q^2}\right>^2}.
\label{eq:Q2_width}
\end{equation}

\begin{figure*}
\centerline{\includegraphics[width=\linewidth]{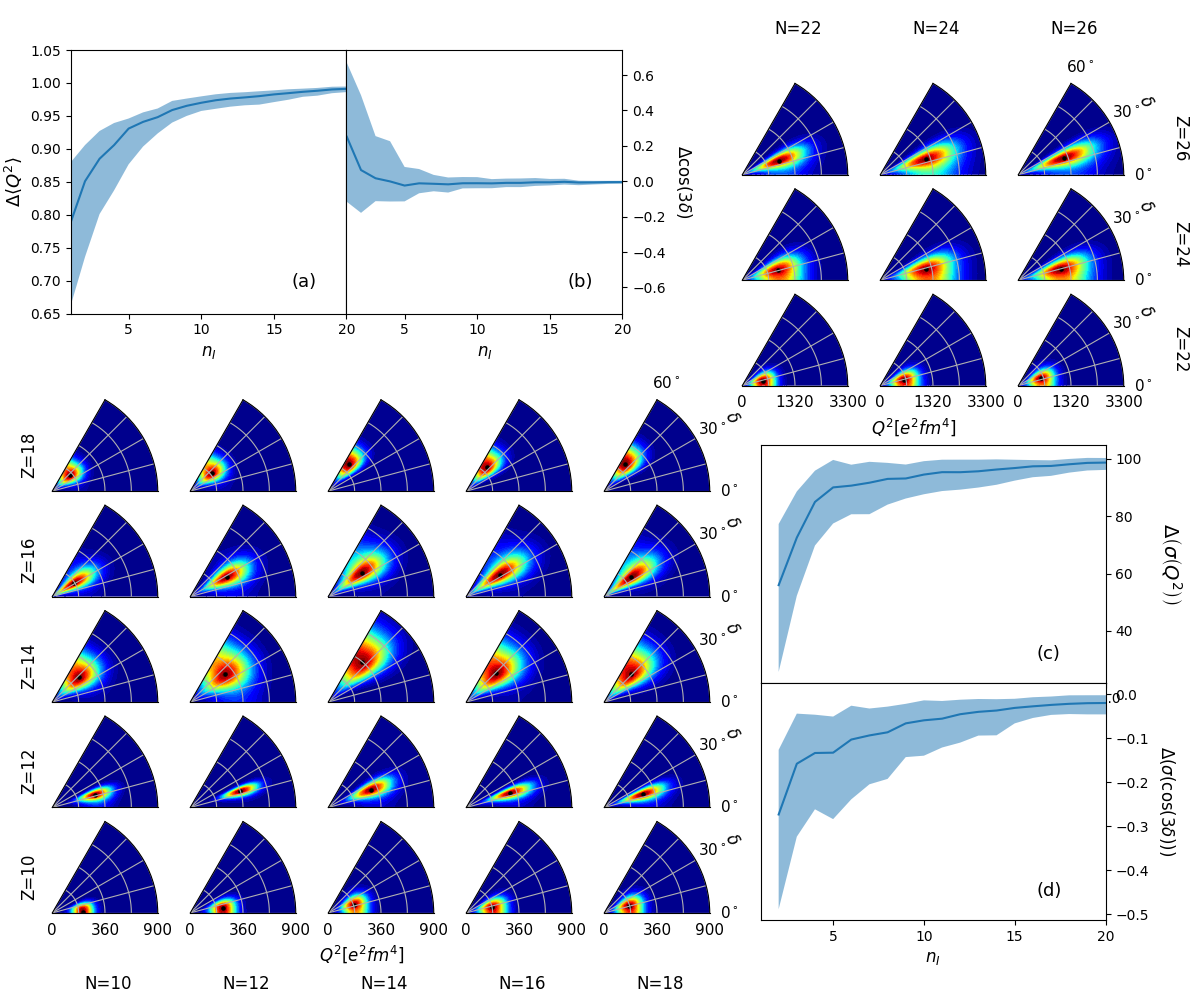}}
\caption{Deformation mapped in $\left(Q^2,\delta\right)$ space for $sd$- and $pf$-shell nuclei as calculated in the present work using nominal effective charges for $n_I=40$. See text for details. Insets: the convergence as a function of the number of shell-model states included in the determination of: (a) $\ev{\hat{Q^2}}$, (b) $\cos{\left(3\delta\right)}$, (c) $\sigma\left(Q^2\right)$ and (d) $\sigma\left(\cos{\left(3\delta\right)}\right)$. The bands indicate the one standard-deviation uncertainties based on the variance of the sample. See~\cref{eq:q2_conv,eq:cos3d_conv,eq:q2sig_conv,eq:cos3dsig_conv}.}
\label{fig:CompFig}
\end{figure*}

The matrix elements contributing to the $\left(J=0\right)$ coupled $\left<\hat{Q^4}\right>$ expectation value in a simplified level scheme are shown in the right-hand panel of Fig.~\ref{fig:loops}, demonstrating the increased requirement for matrix elements. One can also define higher-order invariant quantities, with:

\begin{align}
\left<\hat{P^5}(J)\right>& =  \nonumber \\ &\mel{s}{\{(\hat{E2}\times \hat{E2})_J \times [(\hat{E2} \times \hat{E2})_2 \times \hat{E2}]_J\}_0}{s},
\end{align}

where, 

\begin{multline}
\mel{s}{\{(\hat{E2}\times \hat{E2})_J \times [(\hat{E2} \times \hat{E2})_2 \times \hat{E2}]_J\}_0}{s} \\ 
	= \frac{\sqrt{5}\left(2J+1\right)}{2I_s+1} \sum_{rtvw}M_{st}M_{tr}M_{rv}M_{vw}M_{ws}\\ 
	\left\{\begin{matrix}2 & 2 & J \\ I_s & I_r & I_t \end{matrix}\right\} \left\{\begin{matrix}2 & 2 & J \\ I_s & I_r & I_w \end{matrix}\right\} \left\{\begin{matrix}2 & 2 & 2 \\ I_w & I_r & I_v \end{matrix}\right\} \left(-1\right)^{I_w+I_s}.
\end{multline}

For the $J=0$ case this leads to an expectation value of 

\begin{equation}
\ev{\widehat{Q^5\cos{\left(3\delta\right)}}\left(J=0\right)} = -\sqrt{\frac{35}{2}}\sqrt{5}\ev{\hat{P^5}(0)},
\end{equation}

which provides a secondary measure for $\cos{(3\delta)}$, as well as the possibility to determine the covariance of the first two invariant quantities. Expectation values of $\ev{\hat{Q^6}}$ and $\ev{\widehat{Q^6\cos^2{3\delta}}}$ can also be defined by:

\begin{multline}
\left<\hat{P^6_0}(J)\right> = \\ 
	\bra{s}\{[(\hat{E2}\times\hat{E2})_J\times(\hat{E2}\times\hat{E2})_J]_0\times(\hat{E2}\times\hat{E2})_0\}_0\ket{s}\\ 
\end{multline}

where, 

\begin{equation}
\left<\hat{Q^6}(J=0)\right> = 5\sqrt{5}\left<\hat{P^6_0}(0)\right>,
\end{equation}

yielding

\begin{multline}
	\left<\hat{Q^6}(0)\right> = \\ 
	\frac{5}{2I_s+1}\sum_{\mathclap{\substack{rtvwu\\I_r = I_w}}}\frac{1}{2I_r + 1}M_{st}M_{tw}M_{wu}M_{ur}M_{rv}M_{vs} \\
	\left\{\begin{matrix} 2 & 2 & 0 \\ I_s & I_w & I_t \end{matrix}\right\}\left\{\begin{matrix} 2 & 2 & 0 \\ I_s & I_r & I_v \end{matrix}\right\}  (-1)^{I_s - I_u}.
	\label{eq:P6_0}
\end{multline}

One finally constructs $\left<\hat{P^6}_1(J)\right>$ and $\left<\hat{P^6}_2(J)\right>$ which will be related to $\left<\widehat{Q^6\cos^2{3\delta}}\right>$:

\begin{multline}
\left<\hat{P^6}_1(J)\right> =\\
	 \bra{s}\{[(\hat{E2}\times\hat{E2})_2\times\hat{E2}]_J\times[(\hat{E2}\times\hat{E2})_2\times\hat{E2}]_J\}_0\ket{s}\\ 
	= \frac{5\sqrt{2J+1}}{2I_s + 1}\sum_{rutvw}M_{su}M_{ut}M_{tr}M_{rv}M_{vw}M_{ws} \\
	\left\{\begin{matrix} 2 & 2 & J \\ I_s & I_r & I_t \end{matrix}\right\}\left\{\begin{matrix} 2 & 2 & 2 \\ I_s & I_t & I_u \end{matrix}\right\} \\
	\left\{\begin{matrix} 2 & 2 & J \\ I_s & I_r & I_w \end{matrix}\right\}\left\{\begin{matrix} 2 & 2 & 2 \\ I_w & I_r & I_v \end{matrix}\right\}  (-1)^{2I_s + I_t + I_w}
	\label{eq:P6_1}
\end{multline}

and

\begin{multline}
\left<\hat{P^6}_2(J)\right> = \\
	 \bra{s}\{[(\hat{E2}\times\hat{E2})_2\times\hat{E2}]_J\times[\hat{E2}\times(\hat{E2}\times\hat{E2})_2\}]_J\}_0\ket{s}\\ 
	= \frac{5\sqrt{2J+1}}{2I_s + 1}\sum_{rutvw}M_{su}M_{ut}M_{tr}M_{rv}M_{vw}M_{ws} \\
	\left\{\begin{matrix} 2 & 2 & J \\ I_s & I_r & I_t \end{matrix}\right\}\left\{\begin{matrix} 2 & 2 & 2 \\ I_s & I_t & I_u \end{matrix}\right\} \\
	\left\{\begin{matrix} 2 & 2 & J \\ I_s & I_r & I_v \end{matrix}\right\}\left\{\begin{matrix} 2 & 2 & 2 \\ I_s & I_v & I_w \end{matrix}\right\}  (-1)^{I_s + I_r + I_t + I_w}.
	\label{eq:P6_2}
\end{multline}

From the above quantities one can then determine:

\begin{equation}
\left<\widehat{Q^6\cos^2{3\delta}} (J=0)\right> = \frac{35}{2}\left<\hat{P^6_1}(0)\right> = \frac{35}{2}\left<\hat{P^6_2}(0)\right>.
\end{equation}

The final physical quantity constructed here is the width of the $\cos{3\delta}$ parameter, based on the expectation values calculated above:

\begin{equation}
\sigma(\cos{3\delta}) = \sqrt{\frac{\left<\widehat{Q^6\cos^2{3\delta}}\right>}{\left<\hat{Q^6}\right>} - \left(\frac{\left<\widehat{Q^3\cos{3\delta}}\right>}{\left<\hat{Q^2}\right>^{3/2}}\right)^2}.
\label{eq:tri_width}
\end{equation}

\begin{table*}
\centering
\begin{small}
\begin{tabularx}{\linewidth}{XXXXX}
\hline
		\rule{0pt}{\normalbaselineskip}{\it $n_I$}
 		& 	$\Delta\ev{\hat{Q^2}}$ [\%]
		& 	$\Delta\left(\sigma\left(Q^2\right)\right)$ [\%]
		& 	$\Delta\left(\cos{\left(3\delta\right)}\right)$
		& 	$\Delta\left(\sigma\left(\cos{\left(3\delta\right)}\right)\right)$\\[4pt]
\hline
1		&	$-21.2\pm^{12.6}_{9.2}$	&	Undefined				&	$0.26\pm^{0.42}_{0.37}$	&	Undefined				\\[3pt]
2		&	$-14.9\pm^{11.5}_{5.5}$	&	$-52.4\pm^{19.7}_{24.7}$	&	$0.06\pm^{0.42}_{0.24}$	&	$-0.27\pm^{0.15}_{0.22}$	\\[3pt]
3		&	$-11.5\pm^{8.4}_{4.2}$	&	$-35.5\pm^{14.5}_{20.4}$	&	$0.02\pm^{0.24}_{0.13}$	&	$-0.16\pm^{0.11}_{0.16}$	\\[3pt]
4		&	$-9.4\pm^{6.9}_{3.4}$	&	$-23.1\pm^{10.0}_{14.6}$	&	$0.00\pm^{0.23}_{0.11}$	&	$-0.13\pm^{0.08}_{0.14}$	\\[3pt]
5		&	$-6.9\pm^{5.4}_{1.5}$	&	$-16.2\pm^{9.0}_{12.0}$	&	$-0.02\pm^{0.11}_{0.09}$	&	$-0.13\pm^{0.08}_{0.14}$	\\[3pt]
10		&	$-3.0\pm^{1.2}_{1.0}$	&	$-8.3\pm^{4.7}_{6.0}$	&	$-0.01\pm^{0.03}_{0.03}$	&	$-0.06\pm^{0.04}_{0.07}$	\\[3pt]
15		&	$-1.7\pm^{1.2}_{0.6}$	&	$-4.9\pm^{2.5}_{4.7}$	&	$-0.01\pm^{0.02}_{0.01}$	&	$-0.03\pm^{0.02}_{0.03}$	\\[3pt]
20		&	$-0.9\pm^{0.5}_{0.4}$	&	$-1.8\pm^{1.6}_{2.7}$	&	$0.00\pm^{0.01}_{0.01}$	&	$-0.02\pm^{0.01}_{0.02}$	\\[3pt]
\hline
\end{tabularx}
\end{small}
\caption{Deviation from converged values for $\ev{\hat{Q^2}}$, $\sigma\left(Q^2\right)$, $\cos{\left(3\delta\right)}$ and $\sigma\left(\cos{\left(3\delta\right)}\right)$ using nominal effective charges as defined in~\cref{eq:q2_conv,eq:q2sig_conv,eq:cos3d_conv,eq:cos3dsig_conv} for select values of $n_I$. }
\label{tab:summary}
\vspace{-10pt}
\end{table*}

Note that this definition, and all definitions related to $\cos{(3\delta)}$, assume no covariance between $Q$ and $\delta$. This is a common assumption which will be employed for the majority of the present work, but the role of covariance will be discussed later. An alternate definition of $\sigma\left(\cos{(3\delta)}\right)$ was presented in Ref.~\cite{ref:Poves_19} that includes a covariance term. From the above relations, one can therefore determine an expectation value for the absolute degree of quadrupole deformation $\left<\hat{Q^2}\right>$ (Eq.~\ref{eq:Q2}) and its softness $\sigma\left(Q^2\right)$ (Eq.~\ref{eq:Q2_width}), and the degree of triaxiality $\cos{3\delta}$ (Eq.~\ref{eq:Q3cos3d}) and its softness $\sigma\left(\cos{3\delta}\right)$, (Eq.~\ref{eq:tri_width}). Importantly, these invariant quantities rely only on the spherical tensor nature of the electromagnetic quadrupole operator. Absent any truncation or evolution of the $E2$ operator, one can construct the above invariant quantities from modeled $E2$ matrix elements, as well as experimentally determined values, allowing for like-for-like comparisons. Clearly, the higher-order quantities described in Equations~\ref{eq:P6_0}, ~\ref{eq:P6_1} and \ref{eq:P6_2} require comprehensive sets of matrix elements. This work will take advantage of the model-independence of the determined quantities, along with shell-model calculations in the $sd$- and $pf$-shell model spaces to address the question of convergence: how many experimentally (or theoretically) determined $E2$ matrix elements are required in order to converge on a solution?


\section{Method}

Shell-model calculations were performed with NuShellX@MSU~\cite{ref:NuShellX}. Calculations in the $sd$-shell were performed with the USDB interaction~\cite{ref:USDB} and with effective charges of $e^\pi = 1.36$ and $e^\nu = 0.45$. In the $f_{7/2}$ shell, the full $pf$-model space was used with the KB3G interaction~\cite{ref:KB3G} with effective charges of $e^\pi = 1.5$ and $e^\nu = 0.5$. $E2$ matrix elements determined from the shell-model calculations were then used to calculate the invariant quantities given above. In order to provide a qualitative picture of the results they are presented in a $\left(Q^2,\delta\right)$ space, using Eq.~\ref{eq:cos_approx} to determine $\cos{\left(3\delta\right)}$ and thus $\delta$.  The softness values $\sigma\left(Q^2\right)$ and $\sigma\left(\cos{\left(\delta\right)}\right)$ are treated as standard deviation values within a normal distribution with no asymmetries included. The results can therefore be considered as an approximate probability distribution in $\left(Q^2,\delta\right)$ space, though it is noted that the behaviour of any component of the distribution beyond the limits of $0^\circ\leq\delta\leq60^\circ$ and $Q^2\geq0$ is undefined and as previously mentioned, correlations between $Q$ and $\delta$ are neglected. The calculated ground-state behaviour of all calculated nuclei is shown in Fig.~\ref{fig:CompFig} using all calculated states.

In total, 34 nuclei were calculated which were then used to investigate the convergence of the invariant quantities. As shown in panels (a), (b), (c) and (d) of Fig.~\ref{fig:CompFig}, the number of states of each spin included in the determination ($n_I$) was progressively increased and the difference with the $n_I=40$ values determined, at which point the convergence is assumed to be complete. For example, $n_I=3$ corresponds to the first, second and third $0^+$, $1^+$, $2^+$, etc. states being included in the calculation. In practice though, the Wigner-6j symbols in the above definitions mean that only certain spin states contribute. The convergence was then plotted as follows:

\begin{equation}
\Delta\left<\hat{Q^2}\right> = \left<\hat{Q^2}\right>_{n_I} / \left<\hat{Q^2}\right>_{40} [\%]
\label{eq:q2_conv}
\end{equation}
\begin{equation}
\Delta\cos{\left(3\delta\right)} = \cos{\left(3\delta\right)}_{n_I} - \cos{\left(3\delta\right)}_{40}
\label{eq:cos3d_conv}
\end{equation}
\begin{equation}
\Delta\left(\sigma\left(Q^2\right)\right) =\sigma\left(Q^2\right)_{n_I} / \sigma\left(Q^2\right)_{40} [\%]
\label{eq:q2sig_conv}
\end{equation}
\begin{equation}
\Delta\left(\sigma\left(\cos{\left(3\delta\right)}\right)\right) = \sigma\left(\cos{\left(3\delta\right)}\right)_{n_I} - \sigma\left(\cos{\left(3\delta\right)}\right)_{40}
\label{eq:cos3dsig_conv}
\end{equation}

\begin{figure*}
\centerline{\includegraphics[width=\linewidth]{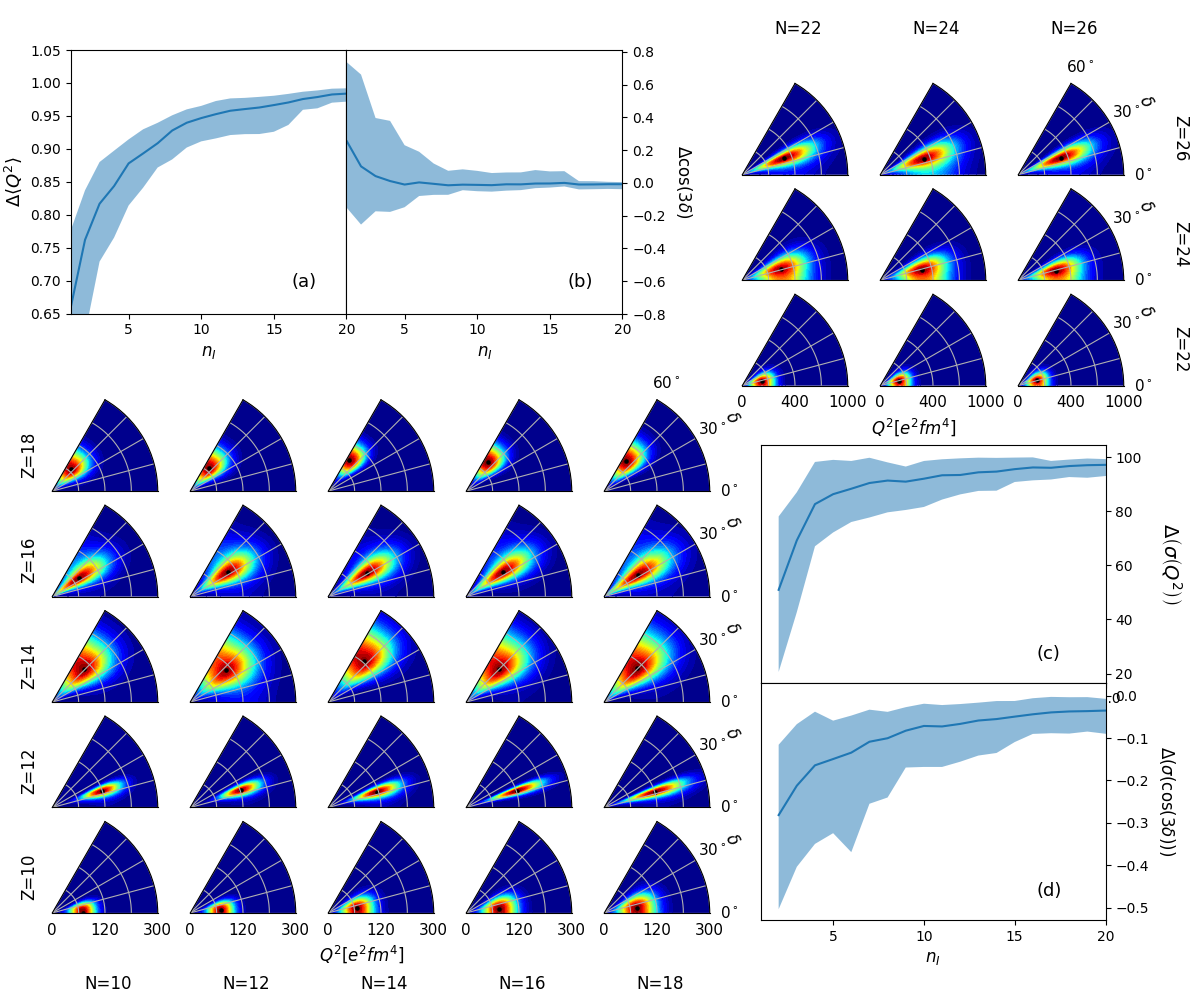}}
\caption{Deformation mapped in $\left(Q^2,\delta\right)$ space for $sd$- and $pf$-shell nuclei as calculated in the present work using bare nucleon charges (i.e. $e^\pi=1$ and $e^\nu=0$) for $n_I=40$. Insets: the convergence as a function of the number of shell-model states included in the determination of: (a) $\ev{\hat{Q^2}}$, (b) $\cos{\left(3\delta\right)}$, (c) $\sigma\left(Q^2\right)$ and (d) $\sigma\left(\cos{\left(3\delta\right)}\right)$. The bands indicate the one standard-deviation uncertainties based on the variance of the sample.  See~\cref{eq:q2_conv,eq:cos3d_conv,eq:q2sig_conv,eq:cos3dsig_conv}.}
\label{fig:CompFig_eff0}
\end{figure*}
\begin{table*}
\centering
\begin{small}
\begin{tabularx}{\linewidth}{XXXXX}
\hline
		\rule{0pt}{\normalbaselineskip}{\it $n_I$}
 		& 	$\Delta\ev{\hat{Q^2}}$ [\%]
		& 	$\Delta\left(\sigma\left(Q^2\right)\right)$ [\%]
		& 	$\Delta\left(\cos{\left(3\delta\right)}\right)$
		& 	$\Delta\left(\sigma\left(\cos{\left(3\delta\right)}\right)\right)$\\[4pt]
\hline
1		&	$-34.2\pm^{21.1}_{11.8}$	&	Undefined				&	$0.26\pm^{0.48}_{0.41}$	&	Undefined				\\[3pt]
2		&	$-23.8\pm^{14.3}_{7.5}$	&	$-62.1\pm^{19.7}_{22.9}$	&	$0.1\pm^{0.56}_{0.35}$	&	$-0.28\pm^{0.17}_{0.22}$	\\[3pt]
3		&	$-18.3\pm^{8.8}_{6.3}$	&	$-43.5\pm^{14.4}_{22.9}$	&	$0.04\pm^{0.35}_{0.21}$	&	$-0.21\pm^{0.14}_{0.19}$	\\[3pt]
4		&	$-15.6\pm^{7.8}_{5.4}$	&	$-30.6\pm^{11.6}_{13.3}$	&	$0.01\pm^{0.37}_{0.19}$	&	$-0.16\pm^{0.13}_{0.19}$	\\[3pt]
5		&	$-12.2\pm^{6.4}_{3.6}$	&	$-24.3\pm^{11.6}_{12.1}$	&	$-0.01\pm^{0.24}_{0.14}$	&	$-0.15\pm^{0.09}_{0.17}$	\\[3pt]
10		&	$-5.3\pm^{3.5}_{1.9}$	&	$-12.8\pm^{6.1}_{10.6}$	&	$-0.01\pm^{0.09}_{0.04}$	&	$-0.07\pm^{0.05}_{0.10}$	\\[3pt]
15		&	$-3.4\pm^{4.0}_{1.4}$	&	$-7.6\pm^{4.0}_{5.5}$	&	$0.00\pm^{0.07}_{0.02}$	&	$-0.05\pm^{0.04}_{0.06}$	\\[3pt]
20		&	$-1.6\pm^{1.2}_{0.8}$	&	$-4.4\pm^{2.5}_{4.1}$	&	$-0.01\pm^{0.01}_{0.02}$	&	$-0.04\pm^{0.03}_{0.05}$	\\[3pt]
\hline
\end{tabularx}
\end{small}
\caption{Deviation from converged values for $\ev{\hat{Q^2}}$, $\sigma\left(Q^2\right)$, $\cos{\left(3\delta\right)}$ and $\sigma\left(\cos{\left(3\delta\right)}\right)$ using bare nucleon charges (i.e. $e^\pi=1$ and $e^\nu=0$) as defined in~\cref{eq:q2_conv,eq:q2sig_conv,eq:cos3d_conv,eq:cos3dsig_conv} for select values of $n_I$.}
\label{tab:summary_eff0}
\vspace{-10pt}
\end{table*}

\begin{figure}
\centerline{\includegraphics[width=.8\linewidth]{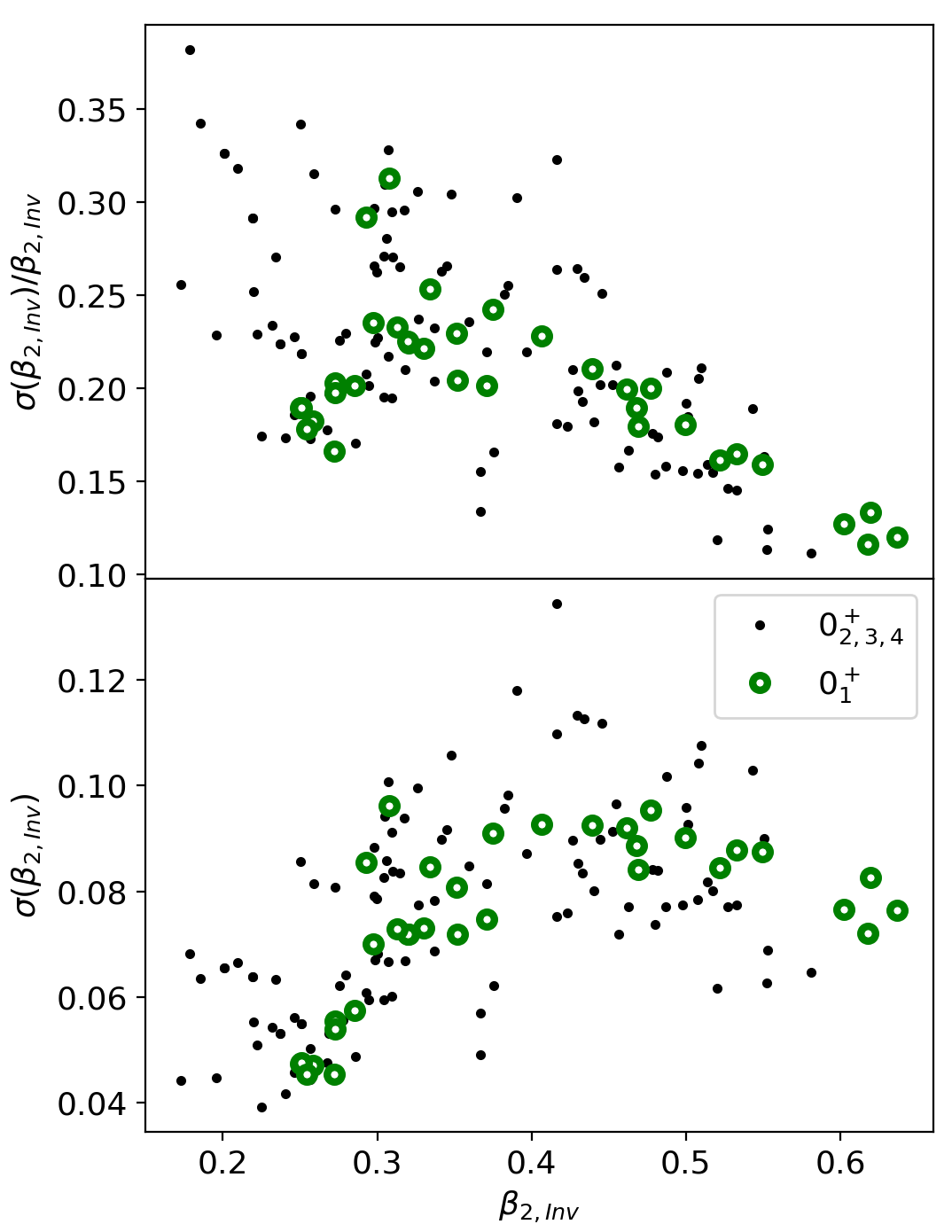}}
\caption{Softness in $\beta_{2,Inv}$ from the ensemble of shell-model data, presented both relative to the $\beta_{2,Inv}$ value (top) and in absolute terms (bottom) and plotted against the deformation parameter $\beta_{2,Inv}$, as defined in Eq.~\ref{eq:beta_2inv} for $n_I=40$. Calculated parameters are shown for both ground and excited $0^+$ states.}
\label{fig:beta}
\end{figure}

\begin{figure}
\centerline{\includegraphics[width=\linewidth]{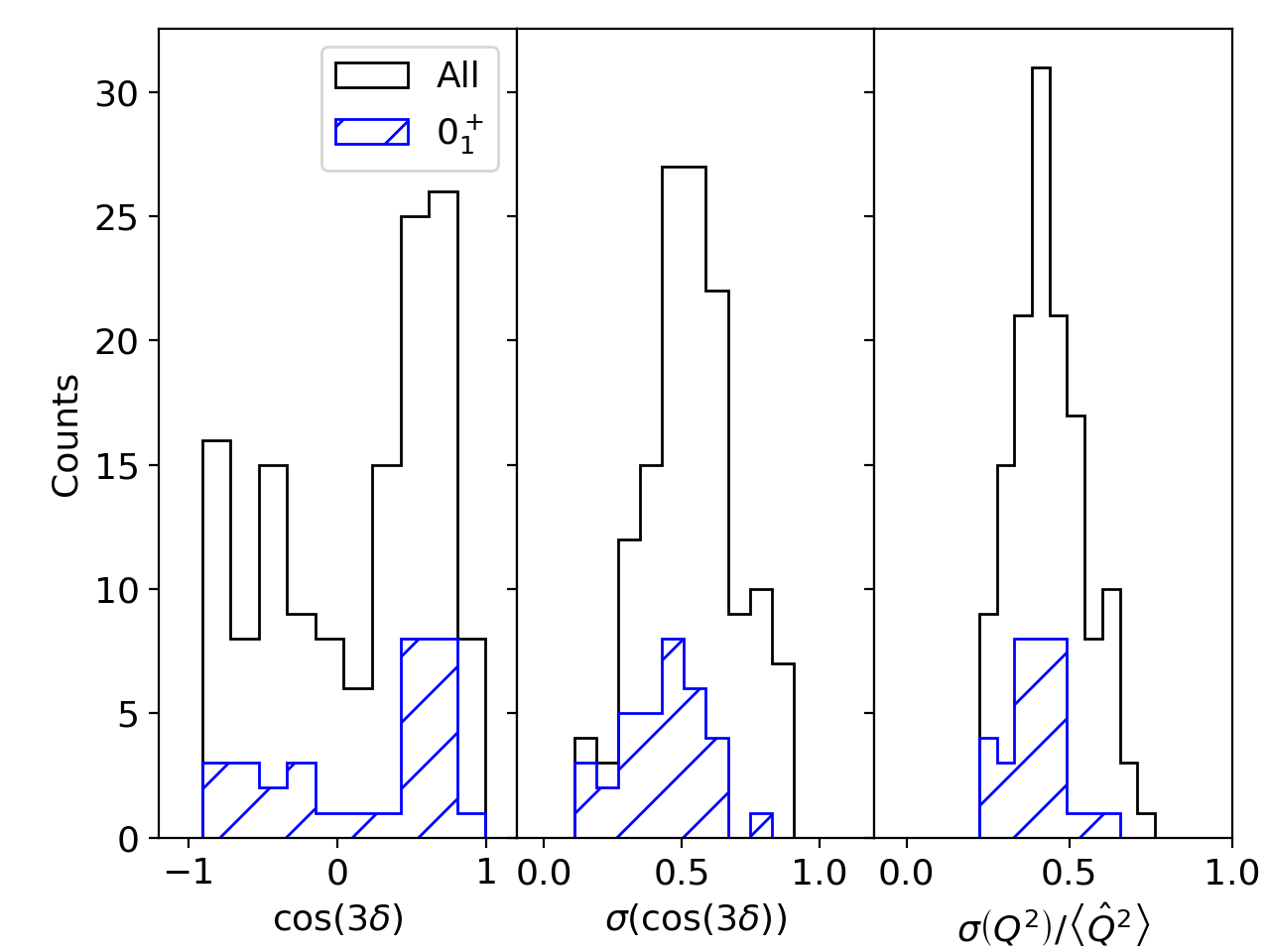}}
\caption{Invariant quantities $\left<\cos{\left(3\delta\right)}\right>$, $\sigma\left(\cos{\left(3\delta\right)}\right)$ and $\sigma\left(Q^2\right)/\left<\hat{Q^2}\right>$ for the full ($n_I=40$) calculations for the ground (filled) and excited $0^+$ states for $n_I=40$.}
\label{fig:hist}
\end{figure}

The assumption of complete convergence by $n_I=40$ is justified by the convergence behaviour of the data ($<1\%$ changes with increasing $n_I$ at $n_I=40$). In addition, where the same nuclei were calculated as in Ref.~\cite{ref:Poves_19} good agreement was found with the $n_I=40$ values. The convergence has no clear dependence on the mass of the nuclei within the model spaces. Qualitatively, one can see that $\ev{\hat{Q^2}}$ and $\cos{\left(3\delta\right)}$ converge rather quickly while the higher-order invariants corresponding to the softness parameters appear slower to converge and have more scatter. Note that, except where explicitly stated otherwise (e.g. when investigating convergence), $\left<\hat{Q^2}\right>$, $cos{\left(3\delta\right)}$, $\sigma{\left(Q^2\right)}$ and $\sigma\left(cos{\left(3\delta\right)}\right)$ values in this work are quoted for $n_I=40$.

One can also treat the data shown in Fig.~\ref{fig:CompFig} as a statistical sample and determine, as a function of $n_I$, the mean deviation from the converged values, as well as the variances of the sample. 
It is found, for example, that the approximate determination of $\ev{\hat{Q^2}}$ in Equation~\ref{eq:q2_LO} is on average deficient from the true value by approximately 20\%, with a standard-deviation of approximately 10\%. Mean deviations and the associated standard deviations on those values for a selection of $n_I$ values are given in Table~\ref{tab:summary}. Note that the invariant quantities required to derive $\left<\hat{Q^4}\right>$, $\left<\hat{Q^6}\right>$ and $\left<\widehat{Q^6cos^2\left(3\delta\right)}\right>$ are undefined for $n_I=1$.

The convergence can also be considered in terms of the shape parameters themselves. While the sample is not sufficient for a quantitative analysis, qualitatively, the nuclei that converge slowly tend to be closer to maximal triaxiality and softer (in both $Q^2$ and $\delta$) than the faster converging cases. 

\section{Discussion}

The underlying assumption of the present work is that the ensemble of nuclei created by the shell-model calculations represents a realistic sample of true atomic nuclei. Importantly it is not essential that the individual nuclei are perfectly reproduced, only that the distribution of nuclei are represented. The role of effective charges, set to nominal values in the above calculations, was investigated. It was found that not including effective charges (i.e. $e^\pi=1$ and $e^\nu=0$) slows the convergence. A summary of the convergence for bare nucleon charges is shown in Fig.~\ref{fig:CompFig_eff0} and Table~\ref{tab:summary_eff0}. 

\subsection{Deformation systematics}

From the present work it is possible to investigate some systematic behaviours of deformation using the ensemble of shell-model data. First, one can redefine Eq.~\ref{eq:beta_2} as

\begin{equation}
\beta_{2,Inv} = \frac{4\pi}{3ZR_0^2}\sqrt{\frac{\ev{\hat{Q^2}}}{e^2}},
\label{eq:beta_2inv}
\end{equation}

with

\begin{equation}
\sigma\left({\beta_{2,Inv}}\right) = \frac{1}{2}\frac{\sigma\left(Q^2\right)}{\ev{\hat{Q^2}}} \beta_{2,Inv}
\end{equation}

which allows for comparisons between nuclei of different masses and proton numbers. Furthermore, the parameters are calculated for the first three excited $0^+$ states in each nucleus, for which sufficient states of higher energy have been calculated to be confident of good convergence. Figure~\ref{fig:beta} shows $\beta_{2,Inv}$ and its softness for the 34 nuclear ground-states and the 102 excited $0^+$ states calculated in this work. A consistent evolution is found for all nuclei with $\sigma\left(\beta_{2,Inv}\right)$, increasing with increasing with $\beta_{2,Inv}$ before reaching a plateau. There is some hint of a reduction in $\sigma\left(\beta_{2,Inv}\right)$ softness occurring beyond $\beta_{2,Inv}\approx0.6$, however the present data are too limited to draw firm conclusions. Final ($n_I=40$) $\cos{\left(3\delta\right)}$, $\sigma\left(\cos{\left(3\delta\right)}\right)$ and $\sigma\left(Q^2\right)/\ev{\hat{Q^2}}$ values are shown in Fig.~\ref{fig:hist}. Notably, the relative softness of $Q^2$ is rather well-localised at about 40\% of the $\ev{\hat{Q^2}}$ expectation value.

A feature that emerges in Fig.~\ref{fig:CompFig_eff0} is the apparent decoupling of proton and neutron shape distributions. Both interactions used are isospin symmetric, meaning that protons (neutrons) in one system behave identically to the neutrons (protons) in the mirror. Noting that the shapes in Fig.~\ref{fig:CompFig_eff0} correspond only to the proton distributions due to the absence of a neutron effective charge, the shape distribution of, for example, \textsuperscript{28}Ar corresponds to the neutron shape in \textsuperscript{28}Ne and vice versa. While the $\ev{\hat{Q^2}}$ values are rather similar in the mirrors, the $\delta$ distributions are in some cases markedly different. The implication is that, at least within the nuclear shell model, the shape of the matter distribution ($\gamma$) does not necessarily correspond closely to that of the charge distribution ($\delta$). This is an important consideration when it comes to comparing calculated matter distributions with experimentally determined electromagnetic transition strengths.

The influence of additional neutrons on the proton shape distribution can also be seen in Fig.~\ref{fig:CompFig_eff0}. Assuming independent proton and neutron distributions, the shapes for a given element would be expected to remain the same. Any change in shape, either in $Q^2$ or in $\delta$ along an isotopic chain is indicative of a change induced by the additional neutrons. In the argon isotopes ($Z=18$), for example, there is a marked increase in $\ev{\hat{Q^2}}$ at $N=14$ and beyond.


\subsection{Shape mixing}

The mixing of different nuclear configurations does influence the $E2$ strength distribution. From a simple perspective, if two bands are mixed then both of the perturbed configurations must be included in order to properly sample the $E2$ strength and the number of states required to achieve convergence must therefore increase. More generally, it is important to note that the rotational invariants sample only the perturbed (mixed) nuclear configurations. This has an important bearing on determining, for example, $\delta$ softness. If two configurations of similar $\ev{\hat{Q^2}}$ but rather different $\delta$ are mixed, the extracted quadrupole invariants presented here will be indistinguishable from a single, $\delta$-soft structure. For example, in Fig.~\ref{fig:CompFig}, \textsuperscript{26}Si is considerably softer in $\delta$ than \textsuperscript{24}Mg. This might be due to a softer intrinsic $\delta$ deformation for a single configuration in \textsuperscript{26}Si, or the mixing of two configurations with individually narrow $\delta$ distributions. \textsuperscript{52}Fe on the other hand exhibits a well defined $\delta$, but with a broad distribution in $Q^2$. This again could be indicative of an inherent softness in $Q^2$, or the mixing of two configurations of comparable $\delta$ but different magnitudes of $Q^2$. Higher-order invariant quantities sampling the third statistical moment (the skew) might provide further insight in this regard but are likely impractical from an experimental perspective, requiring a prohibitive quantity of experimental data.

\subsection{$\beta_2$ determination}

Equation~\ref{eq:beta_2} is commonly used to relate measured $B(E2)$ values to the Bohr $\beta_2$ parameter under the assumption that charge and matter distributions are the same. As discussed in Eq.~\ref{eq:q2_LO}, this relates directly to the approximate (${n_I=1}$) value of $\ev{\hat{Q^2}}$. The analysis presented here demonstrates that the $\ev{\hat{Q^2}}$ value determined from just the inclusion of the $2^+_1\rightarrow0^+_1$ transitions results in approximately 80\% of the final strength. Thus $\beta_2$ determined solely from $B(E2;0^+_1\rightarrow2^+_1)$ values will be similarly deficient by approximately $10\%$.

\subsection{Larger valence spaces}

The present study was necessarily limited to the fairly modest valence spaces of the $sd$- and $fp$-shells. Empirically, it is known that nuclei within larger valence spaces, such as the vast space occupied by rare-Earth nuclei, exhibit properties consistent with those expected of a rigid rotor. A consequence of this behaviour is the suppression of the $\sigma(\delta)$ parameter, as well perhaps as the $\sigma(Q^2)$ value. Generally, this might be expected to result in a reduced fragmentation of the $E2$ strength which might lead to a faster convergence of the invariant sum-rules. Notably, in his original work, Kumar~\cite{ref:Kumar_72} investigated convergence using the available experimental data in \textsuperscript{152}Sm (the first three $2^+$ states) and determined they were sufficient for a ``reasonably good convergence'' of the two leading-order invariant quantities - in approximate agreement with the conclusions of the present work.

\begin{figure*}
\centerline{\includegraphics[width=\linewidth]{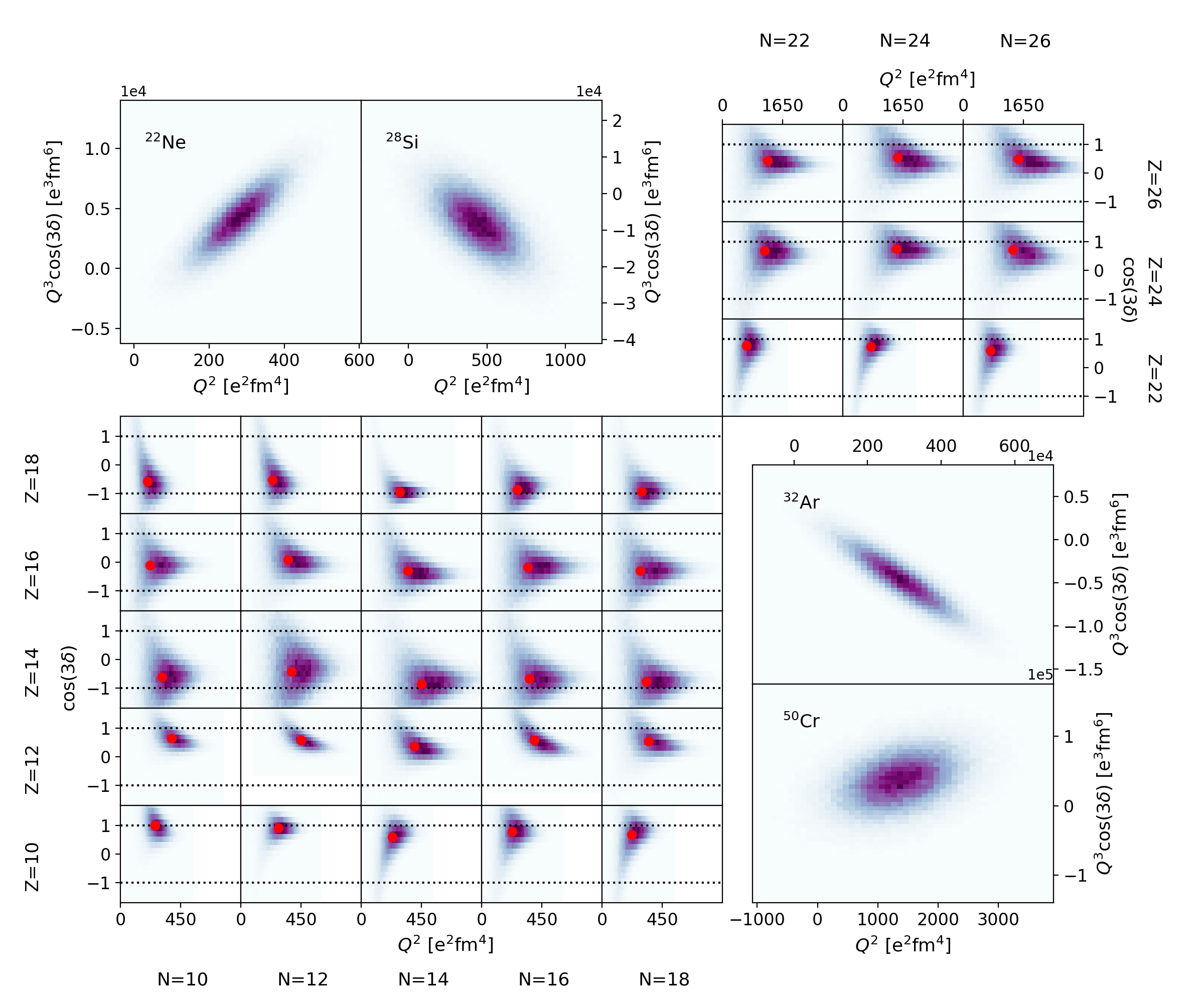}}
\caption{$Q^2$ plotted against $\cos{(3\delta)}$ for each nucleus include in the present study as determined from the covariance matrices for $Q^2$ and $Q^3\cos{(3\delta)}$ calculated using equations~\ref{eq:cov},~\ref{eq:sigq3cos3d} and~\ref{eq:Q2_width} for $n_I=40$. Also shown by the red points are the central values that arise from equations~\ref{eq:Q2} and~\ref{eq:cos} for $n_I=40$. Notably, some fraction of the distribution results in unphysical values, with $\left|\cos{(3\delta)}\right|>1$, as indicated by the dashed lines. Selected examples of the $Q^2$ and $Q^3\cos(3\delta)$ covariance distributions are also shown for $^{22}$Ne, $^{28}$Si, $^{32}$Ar and $^{50}$Cr.}
\label{fig:CompFig_Corr}
\end{figure*}

\subsection{Methodology considerations}

The use of a limited valence space methodology and appropriate interactions will influence the conclusions of the present work. The absence of particle-hole contributions to the wavefunction in this methodology is compensated for by the inclusion of effective charges, which do not alter the wavefunctions themselves. A different framework that incorporates particle-hole excitations might reduce deformation-mixing, which would speed up the convergence seen in Figures~\ref{fig:CompFig} and~\ref{fig:CompFig_eff0}. The present results use a model independent methodology (the sum rules), but the results themselves are model dependent due to the use of a valence space configuration interaction methodology.

\subsection{Parameter correlations}

As mentioned previously, an assumption was made in determining $\cos{3\delta}$ and its softness that $Q$ and $\delta$ are uncorrelated. It is possible to investigate the covariance and correlation of these parameters through the $Q^5\cos{(3\delta)}$ invariant:

\begin{align}
\text{Cov}(Q^2,Q^3\cos{(3\delta)})& = \\ \nonumber  &\ev{Q^5\cos{(3\delta)}} - \ev{Q^2}\ev{Q^3\cos{(3\delta)}} \label{eq:cov} \\
\text{Corr}(Q^2,Q^3\cos{(3\delta)})& = \\ \nonumber  &\frac{\ev{Q^5\cos{(3\delta)}} - \ev{Q^2}\ev{Q^3\cos{(3\delta)}}}{\sigma(Q^2)\sigma(Q^3\cos{(3\delta)}},
\end{align}

where $\sigma(Q^3\cos{(3\delta)})$ is derived as

\begin{equation}
\sigma(Q^3\cos{(3\delta)}) = \sqrt{\ev{Q^6\cos^2{(3\delta)}} - \ev{Q^3\cos{(3\delta)}}^2}.
\label{eq:sigq3cos3d}
\end{equation}

\noindent The joint distribution of $Q^2$ and $Q^3\cos(3\delta)$ is treated using a bivariate normal distribution and incorporating the covariance and variance of the parameters, which is then sampled. For each sample of the distribution, $\cos{(3\delta)}$ is calculated as in Eq.~\ref{eq:cos}. Figure~\ref{fig:CompFig_Corr} shows the result of this analysis on $Q^2$ and $\cos{(3\delta)}$ for all of the nuclei in the present study, with selected examples of the $Q^2$ and $Q^3\cos(3\delta)$ correlation.

The nuclei in Fig.~\ref{fig:CompFig_Corr} exhibit varying degrees of correlation between $Q^2$ and $\cos{(3\delta)}$. The uncorrelated uncertainties included in Figs.~\ref{fig:CompFig} and ~\ref{fig:CompFig_eff0} therefore constitute a lower bound on the breadth of the distribution in $\delta$, although the effect is typically not dramatic. Notably a significant component of the probability distribution exists in an unphysical regime ($\left|\cos{(3\delta)}\right|>1$). In part this is due to the use of a simple bivariate normal distribution to describe the probability distribution. A more sophisticated parameterization which properly incorporates the full covariance matrix as well as the physical limits may help to overcome this, but is beyond the scope of the present work.

\subsection{Future prospects}

The use of the nuclear shell model for the present work has many advantages. The nuclear structures created in the $sd$- and lower $pf$-shell model spaces are well reproduced experimentally at low energies and the calculations begin with no assumptions about the nuclear shape which might otherwise bias the results. A similar analysis using different theoretical methods would be invaluable however, in particular to provide a more global picture: realistically the present analysis is limited to nuclei with $A\lesssim70$. Experimentally, identifying cases in which comprehensive experimental data can be collected for the first five $2^+$ states, for example, would provide some confirmation of the convergent behaviour demonstrated here. Such experimental measurements might reasonably be achieved in the heavier mass regions with state-of-the-art experimental equipment and high-intensity stable beams, such as employed in Ref.~\cite{ref:Ayangeakaa_16,ref:Ayangeakaa_19}.

\section{Conclusions}

The nuclear shell model has been employed in the $sd$- and $pf$-shell model spaces to investigate the convergence of the quadrupole sum-rules~\cite{ref:Kumar_72,ref:Cline_86}. Large numbers of nuclear states were calculated, allowing for progressively more comprehensive sets of $E2$ matrix elements to be used in the sum-rule determination. Treating the 34 nuclei as a statistical sample, mean convergence properties were deduced, along with standard deviations. While sensitivity requirements will vary on a case by case basis, it is found that the mean values, $\ev{\hat{Q^2}}$ and $\cos{3\delta}$ converge rapidly. On average, by $n_I$=4, $\ev{\hat{Q^2}}$ has converged to better than 10\% of its true value, while $\cos{3\delta}$ has converged to approximately its true value. While this average convergence holds, there remains some significant scatter about the average values. Higher-order invariant quantities relating to the softness of these values require more data, with $n_I$=5 arguably required to draw any strong conclusions with regards to softness. These results have significant importance for the determination of $\beta$ and $\gamma$ softness in nuclei, in particular with regards to the search for truly rigid structures, where a well-converged $Q$ and $\delta$ softness is required.

\section{Acknowledgements}

The work at LLNL is under contract DE-AC52-07NA27344. Discussions with S.~R.~Stroberg, P.~Adsley, B.~A.~Brown, J.~M.~Allmond, L.~P.~Gaffney, A.~Poves, G.~L.~Wilson, and C.~Y.~Wu are gratefully acknowledged.

\bibliographystyle{unsrt}
\bibliography{invariant}

\begin{thebibliography}{10}

\bibitem{ref:Kumar_72}
K.~Kumar.
\newblock {\em Physical Review Letters}, 28:249, 1972.

\bibitem{ref:Cline_86}
D.~Cline.
\newblock {\em Annual Review of Nuclear and Particle Science}, 36:681, 1986.

\bibitem{ref:Ayangeakaa_19}
A.~D. Ayangeakaa, R.~V.~F. Janssens, S.~Zhu, D.~Little, J.~Henderson, C.~Y. Wu,
  D.~J. Hartley, M.~Albers, K.~Auranen, B.~Bucher, M.~P. Carpenter,
  P.~Chowdhury, D.~Cline, H.~L. Crawford, P.~Fallon, A.~M. Forney, A.~Gade,
  A.~B. Hayes, F.~G. Kondev, Krishichayan, T.~Lauritsen, J.~Li, A.~O.
  Macchiavelli, D.~Rhodes, D.~Seweryniak, S.~M. Stolze, W.~B. Walters, and
  J.~Wu.
\newblock {\em Phys. Rev. Lett.}, 123:102501, 2019.

\bibitem{ref:Henderson_19}
J.~Henderson, C.~Y. Wu, J.~Ash, B.~A. Brown, P.~C. Bender, R.~Elder, B.~Elman,
  A.~Gade, M.~Grinder, H.~Iwasaki, B.~Longfellow,
  T.~Mijatovi\ifmmode~\acute{c}\else \'{c}\fi{}, D.~Rhodes, M.~Spieker, and
  D.~Weisshaar.
\newblock Triaxiality in selenium-76.
\newblock {\em Phys. Rev. C}, 99:054313, 2019.

\bibitem{ref:Ayangeakaa_16}
A.D. Ayangeakaa, R.V.F. Janssens, C.Y. Wu, J.M. Allmond, J.L. Wood, S.~Zhu,
  M.~Albers, S.~Almaraz-Calderon, B.~Bucher, M.P. Carpenter, C.J. Chiara,
  D.~Cline, H.L. Crawford, H.M. David, J.~Harker, A.B. Hayes, C.R. Hoffman,
  B.P. Kay, K.~Kolos, A.~Korichi, T.~Lauritsen, A.O. Macchiavelli, A.~Richard,
  D.~Seweryniak, and A.~Wiens.
\newblock {\em Physics Letters B}, 754:254 -- 259, 2016.

\bibitem{ref:Hadynska_16}
K.~Hady\ifmmode \acute{n}\else \'{n}\fi{}ska-Kl\ifmmode~\mbox{\c{e}}\else
  \c{e}\fi{}k, P.~J. Napiorkowski, M.~Zieli\ifmmode~\acute{n}\else
  \'{n}\fi{}ska, J.~Srebrny, A.~Maj, F.~Azaiez, J.~J. Valiente~Dob\'on,
  M.~Kici\ifmmode \acute{n}\else~\'{n}\fi{}ska Habior, F.~Nowacki,
  H.~Na\"{\i}dja, B.~Bounthong, T.~R. Rodr\'{\i}guez, G.~de~Angelis,
  T.~Abraham, G.~Anil~Kumar, D.~Bazzacco, M.~Bellato, D.~Bortolato,
  P.~Bednarczyk, G.~Benzoni, L.~Berti, B.~Birkenbach, B.~Bruyneel,
  S.~Brambilla, F.~Camera, J.~Chavas, B.~Cederwall, L.~Charles, M.~Ciema\l{}a,
  P.~Cocconi, P.~Coleman-Smith, A.~Colombo, A.~Corsi, F.~C.~L. Crespi, D.~M.
  Cullen, A.~Czermak, P.~D\'esesquelles, D.~T. Doherty, B.~Dulny, J.~Eberth,
  E.~Farnea, B.~Fornal, S.~Franchoo, A.~Gadea, A.~Giaz, A.~Gottardo, X.~Grave,
  J.~Gr\ifmmode~\mbox{\c{e}}\else \c{e}\fi{}bosz, A.~G\"orgen, M.~Gulmini,
  T.~Habermann, H.~Hess, R.~Isocrate, J.~Iwanicki, G.~Jaworski, D.~S. Judson,
  A.~Jungclaus, N.~Karkour, M.~Kmiecik, D.~Karpi\ifmmode~\acute{n}\else
  \'{n}\fi{}ski, M.~Kisieli\ifmmode~\acute{n}\else \'{n}\fi{}ski,
  N.~Kondratyev, A.~Korichi, M.~Komorowska, M.~Kowalczyk, W.~Korten,
  M.~Krzysiek, G.~Lehaut, S.~Leoni, J.~Ljungvall, A.~Lopez-Martens, S.~Lunardi,
  G.~Maron, K.~Mazurek, R.~Menegazzo, D.~Mengoni, E.~Merch\'an, W.~M\ifmmode
  \mbox{\c{e}}\else \c{e}\fi{}czy\ifmmode~\acute{n}\else \'{n}\fi{}ski,
  C.~Michelagnoli, J.~Mierzejewski, B.~Million, S.~Myalski, D.~R. Napoli,
  R.~Nicolini, M.~Niikura, A.~Obertelli, S.~F. \"Ozmen, M.~Palacz,
  L.~Pr\'ochniak, A.~Pullia, B.~Quintana, G.~Rampazzo, F.~Recchia, N.~Redon,
  P.~Reiter, D.~Rosso, K.~Rusek, E.~Sahin, M.-D. Salsac, P.-A. S\"oderstr\"om,
  I.~Stefan, O.~St\'ezowski, J.~Stycze\ifmmode~\acute{n}\else \'{n}\fi{}, Ch.
  Theisen, N.~Toniolo, C.~A. Ur, V.~Vandone, R.~Wadsworth, B.~Wasilewska,
  A.~Wiens, J.~L. Wood, K.~Wrzosek-Lipska, and M.~Zi\ifmmode \mbox{\c{e}}\else
  \c{e}\fi{}bli\ifmmode~\acute{n}\else \'{n}\fi{}ski.
\newblock {\em Phys. Rev. Lett.}, 117:062501, 2016.

\bibitem{ref:Wrzosek_12}
K.~Wrzosek-Lipska, L.~Pr\'ochniak, M.~Zieli\'{n}ska, J.~Srebrny,
  K.~Hady\'{n}ska-Klek, J.~Iwanicki, M.~Kisieli\'{n}ski, M.~Kowalczyk, P.~J.
  Napiorkowski, D.~Pietak, and T.~Czosnyka.
\newblock {\em Phys. Rev. C}, 86:064305, 2012.

\bibitem{ref:Clement_07}
E.~Cl\'ement, A.~G\"orgen, W.~Korten, E.~Bouchez, A.~Chatillon, J.-P.
  Delaroche, M.~Girod, H.~Goutte, A.~H\"urstel, Y.~Le Coz, A.~Obertelli,
  S.~P\'eru, Ch. Theisen, J.~N. Wilson, M.~Zieli\ifmmode~\acute{n}\else
  \'{n}\fi{}ska, C.~Andreoiu, F.~Becker, P.~A. Butler, J.~M. Casandjian, W.~N.
  Catford, T.~Czosnyka, G.~de France, J.~Gerl, R.-D. Herzberg, J.~Iwanicki,
  D.~G. Jenkins, G.~D. Jones, P.~J. Napiorkowski, G.~Sletten, and C.~N. Timis.
\newblock {\em Phys. Rev. C}, 75:054313, 2007.

\bibitem{ref:Srebny_06}
J.~Srebrny, T.~Czosnyka, Ch. Droste, S.G. Rohozi{\'n}ski, L.~Pr{\'o}chniak,
  K.~Zaj\'{i}c, K.~Pomorski, D.~Cline, C.Y. Wu, A.~B{\"a}cklin, L.~Hasselgren,
  R.M. Diamond, D.~Habs, H.J. K{\"o}rner, F.S. Stephens, C.~Baktash, and R.P.
  Kostecki.
\newblock {\em Nuclear Physics A}, 766:25 -- 51, 2006.

\bibitem{ref:Wu_96}
C.Y. Wu, D.~Cline, T.~Czosnyka, A.~Backlin, C.~Baktash, R.M. Diamond, G.D.
  Dracoulis, L.~Hasselgren, H.~Kluge, B.~Kotlinski, J.R. Leigh, J.O. Newton,
  W.R. Phillips, S.H. Sie, J.~Srebrny, and F.S. Stephens.
\newblock {\em Nuclear Physics A}, 607:178 -- 234, 1996.

\bibitem{ref:Poves_19}
A.~Poves, F.~Nowacki, and Y.~Alhassid.
\newblock {\em Phys. Rev. C}, 101:054307, 2020.

\bibitem{ref:Gilbreth_18}
C.~N. Gilbreth, Y.~Alhassid, and G.~F. Bertsch.
\newblock {\em Phys. Rev. C}, 97:014315, 2018.

\bibitem{ref:Naidja_17}
H.~Na\"{\i}dja, F.~Nowacki, and B.~Bounthong.
\newblock {\em Phys. Rev. C}, 96:034312, 2017.

\bibitem{ref:Quan_17}
S.~Quan, Q.~Chen, Z.~P. Li, T.~Nik\ifmmode \check{s}\else
  \v{s}\fi{}i\ifmmode~\acute{c}\else \'{c}\fi{}, and D.~Vretenar.
\newblock {\em Phys. Rev. C}, 95:054321, 2017.

\bibitem{ref:Schmidt_17}
T.~Schmidt, K.~L.~G. Heyde, A.~Blazhev, and J.~Jolie.
\newblock {\em Phys. Rev. C}, 96:014302, 2017.

\bibitem{ref:Pritychenko_16}
B.~Pritychenko, M.~Birch, B.~Singh, and M.~Horoi.
\newblock {\em Atomic Data and Nuclear Data Tables}, 107:1, 2016.

\bibitem{ref:Davydov_58}
A.S. Davydov and G.F. Filippov.
\newblock Rotational states in even atomic nuclei.
\newblock {\em Nuclear Physics}, 8:237 -- 249, 1958.

\bibitem{ref:GOSIA_manual}
D.~Cline, T.~Czosnyka, A.~B. Hayes, P.~Napiorkowski, N.~Warr, and C.~Y. Wu.
\newblock Gosia manual.
\newblock Version: May 10, 2012.

\bibitem{ref:Alhassid_14}
Y.~Alhassid, C.~N. Gilbreth, and G.~F. Bertsch.
\newblock {\em Physical Review Letters}, 113:262503, 2014.

\bibitem{ref:NuShellX}
B.~A. Brown and W.~D.~M. Rae.
\newblock {\em Nucl. Data. Sheets}, 120:115, 2014.

\bibitem{ref:USDB}
B.~A. Brown and W.~A. Richter.
\newblock {\em Phys. Rev. C}, 74:034315, 2006.

\bibitem{ref:KB3G}
A.~Poves, J.~S\'anchez-Solano, E.~Caurier, and F.~Nowacki.
\newblock {\em Nuclear Physics A}, 694:157, 2001.

\end{thebibliography}

\end{document}